\documentclass[aps,prb,twocolumn,superscriptaddress,showpacs,floatfix,longbibliography]{revtex4-1}
\usepackage{graphicx}
\usepackage{dcolumn}
\usepackage{bm}

\usepackage{multirow}

\usepackage{amssymb}
\usepackage{amsmath}
\usepackage{graphicx}
\usepackage[colorlinks,citecolor=blue,linkcolor=red,urlcolor=blue]{hyperref}

\begin{document}
\title{Nonequilibrium dynamics in deconfined quantum critical point revealed by imaginary-time evolution}
\author{Yu-Rong Shu}
\affiliation{School of Physics and Electronic Engineering, Guangzhou University, Guangzhou 510006, China}
\affiliation{Research Center for Advanced Information Materials, Guangzhou University, Guangzhou 510006, China}
\author{Shao-Kai Jian}
\affiliation{Condensed Matter Theory Center, Department of Physics, University of Maryland, College Park, Maryland 20742, USA}
\author{Shuai Yin}
\email{yinsh6@mail.sysu.edu.cn}
\affiliation{School of Physics, Sun Yat-Sen University, Guangzhou 510275, China}
\date{\today}

\begin{abstract}

\end{abstract}

\maketitle

{\bf As proposed to describe putative continuous phase transitions between two ordered phases, the deconfined quantum critical point (DQCP) goes beyond the prevalent Landau-Ginzburg-Wilson (LGW) paradigm since its critical theory is not expressed in terms of the order parameters characterizing either state, but involves fractionalized degrees of freedom and an emergent symmetry~\cite{Senthil2004,Senthil2004b}. So far, great efforts have been spent on its equilibrium properties~\cite{Senthil2004,Senthil2004b,Sandvik2007,Melko2008,Jiang2008,Kuklov2008,Lou2009,Sandvik2010,Chen2013,Harada2013,Pujari2013,Nahum2015,Shao2016,Zayed2017,Wang2017,Ihrig2019}, but the nonequilibrium properties therein are largely unknown. Here we study the nonequilibrium dynamics of the DQCP via the imaginary-time evolution in the two-dimensional (2D) J-Q$_3$ model. We discover fascinating nonequilibrium scaling behaviors hinging on the process of fractionization and the dynamics of emergent symmetry associated with two length scales. Our findings not only constitute a new realm of nonequilibrium criticality in DQCP, but also offer a controllable knob by which to investigate the dynamics in strongly correlated systems.}

Exotic phenomena often emerge in the $2$D quantum magnetic systems~\cite{Sachdev2008}. A prominent example is the possible Landau-forbidden continuous phase transition between the N\'{e}el order and the spontaneously dimerized valence-bond solid (VBS) in the spin-$1/2$ Heisenberg model. To explain it, the DQCP theory was proposed by showing that the dominant fluctuating modes near this critical point are the deconfined spinons and the emergent gauge fields, and both the N\'{e}el and the VBS order parameters are composites of these deconfined objects, rather than fundamental variables~\cite{Senthil2004,Senthil2004b}. To reconcile these two orders with non-compatible broken symmetries, an emergent SO($5$) symmetry appears at the critical point~\cite{Nahum2015b}, demonstrating that the critical theory of DQCP can be cast into the noncompact CP$^1$ model with the monopole fugacity as its dangerously irrelevant scaling variable~\cite{Senthil2004,Senthil2004b}. Pertinent to this, an extra divergent length $\xi'$, which measures the spinon confinement length or the thickness of the VBS domain walls, develops, in addition to the conventional correlation length $\xi$~\cite{Senthil2004b,Shao2016}. And they satisfy $\xi'\propto \xi^{\frac{\nu'}{\nu}}$ with $\nu$ and $\nu'$ being the corresponding critical exponents~\cite{Senthil2004b,Shao2016}. It was plausibly shown that the interplay between these two length scales may take responsibility for the anomalous equilibrium scaling behaviors near the DQCP~\cite{Nahum2015,Shao2016}. 

Near a critical point, the nonequilibrium dynamics is uniquely essential due to the critical slowing down. Extensive studies have been invested in the nonequilibrium behaviors in both classical and quantum phase transitions~\cite{Dziarmaga2010,Polkovnikov2011}. For the DQCP, fundamental and interesting questions arise: How does the deconfined dynamic process happen at the critical point? Is there any specific scaling behavior associated with this fractionalized procedure? How do the two length scales affect the nonequilibrium critical dynamics?

To answer these questions, we explore the nonequilibrium dynamics of the DQCP in imaginary-time direction. It was shown that the imaginary-time dynamics not only shares some universal properties with the real-time dynamics~\cite{DeGrandi2011}, but also bears amenability to large-scale quantum Monte Carlo (QMC) simulations without sign-problem~\cite{DeGrandi2011}. Besides, the imaginary-time evolution recently finds its application in quantum computers~\cite{Motta2020}. Moreover, previous studies demonstrated that the short-imaginary-time dynamics (SITD) in the LGW quantum phase transitions~\cite{Yin2014} exhibits scaling behaviors in analogy to the classical short-time critical dynamics~\cite{Janssen1989,Li1996,Albano2011}, providing fruitful insights in quantum critical dynamics~\cite{Yin2014}.

\begin{figure*}[htbp]
\centering
  \includegraphics[width=\linewidth,clip]{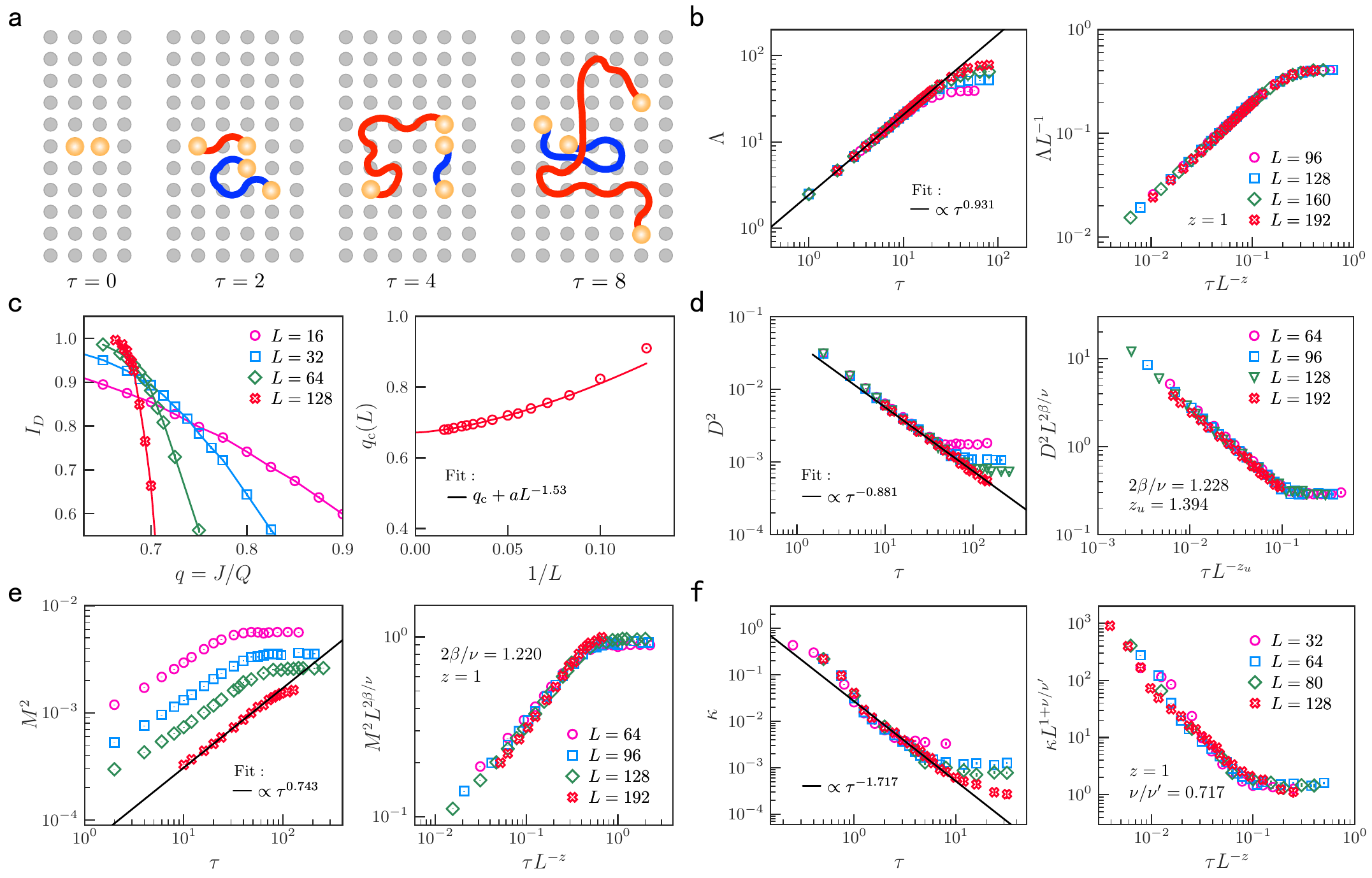}
  \vskip-3mm
  \caption{\textbf{Relaxation dynamics with the VBS initial state.} \textbf{a,} Illustration of the fractionalization process of spinons. Shown is the evolution of typical configurations for a sampled overlap $\langle\psi_{\textrm{left}}|\psi_{\textrm{right}}\rangle$ in $S=1$ sector with two spinon strings which are initially locate in nearest-neighbor sites embedded in the VBS background (not shown). \textbf{b,} At the critical point, in the short-time stage, the size of spinon pair, $\Lambda$, increases with the imaginary-time $\tau$ as $\Lambda\propto \tau^{0.931}$, obtained by power fitting. This exponent is close to $1$, indicating that $\Lambda\propto \tau^{1/z}$. This result is further confirmed by the finite-size scaling (FSS) collapse according to $\Lambda=Lf(\tau L^{-z})$ as shown in \textbf{b} (right). \textbf{c,} The average sign of the VBS order parameter $I_D$ is used to determine the critical point. For different $L$ and fixed $\tau L^{-1}=1/4$, crossing points (left) of curves of $I_D-q$ are extrapolated to estimate the critical point as $q_c\simeq0.671(2)$ (right), which is consistent with the known results~\cite{Lou2009}. \textbf{d,} At $q=q_c$, in the long-time stage, $D^2\propto L^{-1.228}$ (Supplementary Information), giving that $2 \beta/\nu\simeq1.228(4)$, consistent with the known results~\cite{Lou2009,Harada2013,Nahum2015b}. In the short-time stage, $D^2$ decays as $D^2\propto\tau^{-0.881}$ (left), suggesting the exponent is $2 \beta/\nu z_u\simeq0.881(11)$ with $z_u\equiv z \nu'/\nu\simeq 1.394$. Accordingly, $\nu/\nu'\simeq0.717$, close to the known results~\cite{Shao2016}. For convenience, the scaling form $D^2(\tau,L)=\tau^{-2\beta/\nu z_u}f(\tau L^{-z_u})$ is reshaped as $D^2(\tau,L)=L^{-2\beta/\nu}f(\tau L^{-z_u})$ by a simple replacement of the scaling variable (Similar replacements are also implemented for other quantities as follows). Then the latter is examined by the FSS collapse as shown in \textbf{d} (right). \textbf{e,} In the short-time region, $M^2$ increases as $M^2\propto \tau^{0.743}$ (left), with its exponent close to ${(d-2\beta/\nu)/z}$, and depends on $L$ as $M^2\propto L^{-2}$ (Supplementary Information). The scaling form $M^2(\tau,L)=L^{-d}\tau^{d/z-2\beta/\nu z}f(\tau L^{-z})$ is then verified (right). \textbf{f,} The density of domain wall energy $\kappa$ decays as $\kappa\propto\tau^{-1/z-1/z_u}$ (left) in the short-time region and its full relaxation process satisfies $\kappa(\tau,L)=\tau^{-1/z-1/z_u}f(\tau L^{-z})$, verified by FSS (right). These results show that at the critical point $I_D$ and $M^2$ is controlled by the dynamics of $\xi'$, $D^2$ is controlled by the dynamics of $\xi$, and $\kappa$ is controlled by both.
  }
  \label{fig:vbs}
\end{figure*}

Here we sudy the SITD in DQCP.  The J-Q$_3$ model is taken as an example~\cite{Sandvik2007}. Its Hamiltonian reads
\begin{equation}
\label{eq:hamiltonian}
H=-J\sum_{\langle ij\rangle}P_{ij}-Q\sum_{\langle ijklmn\rangle}{P_{ij}P_{kl}P_{mn}},
\end{equation}
in which $J>0$ and $Q>0$, $\langle ij\rangle$ and $\langle ijklmn\rangle$ denote, respectively, nearest neighbors and three nearest-neighbor pairs in horizontal or vertical columns on the square lattice, and $P_{ij}$ denotes the spin singlet operator defined as $P_{ij}\equiv\frac{1}{4}-{\bf S}_{i}\cdot{\bf S}_{j}$ with $\bf S$ being the spin-$1/2$ operator. The system favors the N\'{e}el (VBS) phase with a finite order parameter $M$ ($D$) when $q\equiv J/Q\gg 1$ ($\ll1$). For the SITD, the evolution of the wave function $|\psi(\tau)\rangle$ obeys the imaginary-time Schr\"{o}dinger equation $-\frac{\partial}{\partial\tau}|\psi(\tau)\rangle=H|\psi(\tau)\rangle$ with an uncorrelated initial state~\cite{Yin2014}.

We first study the dynamics of two length scales. The SITD theory in the LGW phase transition with a single length scale $\xi$ shows that for an uncorrelated initial state, $\xi$ increases with $\tau$ as $\xi\propto \tau^{\frac{1}{z}}$ with $z$ the dynamic exponent~\cite{Yin2014}. In the DQCP it is widely believed that $z=1$~\cite{Senthil2004b} and thus one might take it for granted that $\xi\propto \tau^{\frac{1}{z}}$ and $\xi'\propto \tau^{\frac{\nu'}{\nu z}}$.

Surprisingly, this assumption is not the case. In contrast, we identify that it is the spinon confinement length $\xi'$ rather than the usual correlation length $\xi$ that increases with $\tau$ as $\tau^{\frac{1}{z}}$. To see this, we study the deconfined process from an initial state with a triplet embedded in the VBS background at the critical point (determined in the following). We find that the size of the spinon pair $\Lambda$, defined via the strings connecting the unpaired spins in the $S=1$ sector~\cite{Shao2016,Tang2013}, increases with $\tau$ as $\Lambda\propto \tau^{\frac{1}{z}}$ as shown in Fig.~\ref{fig:vbs}~{\bf a,b}. This demonstrates that $\xi'\propto \tau^{\frac{1}{z}}$ since it was shown that $\xi'\propto \Lambda$~\cite{Shao2016}. There follows $\xi\propto \tau^{\frac{1}{z_u}}$ with $z_u\equiv \frac{z \nu'}{\nu}$ (subscript $u$ indicates usual length scale).
\begin{figure*}[htbp]
\centering
  \includegraphics[width=\linewidth,clip]{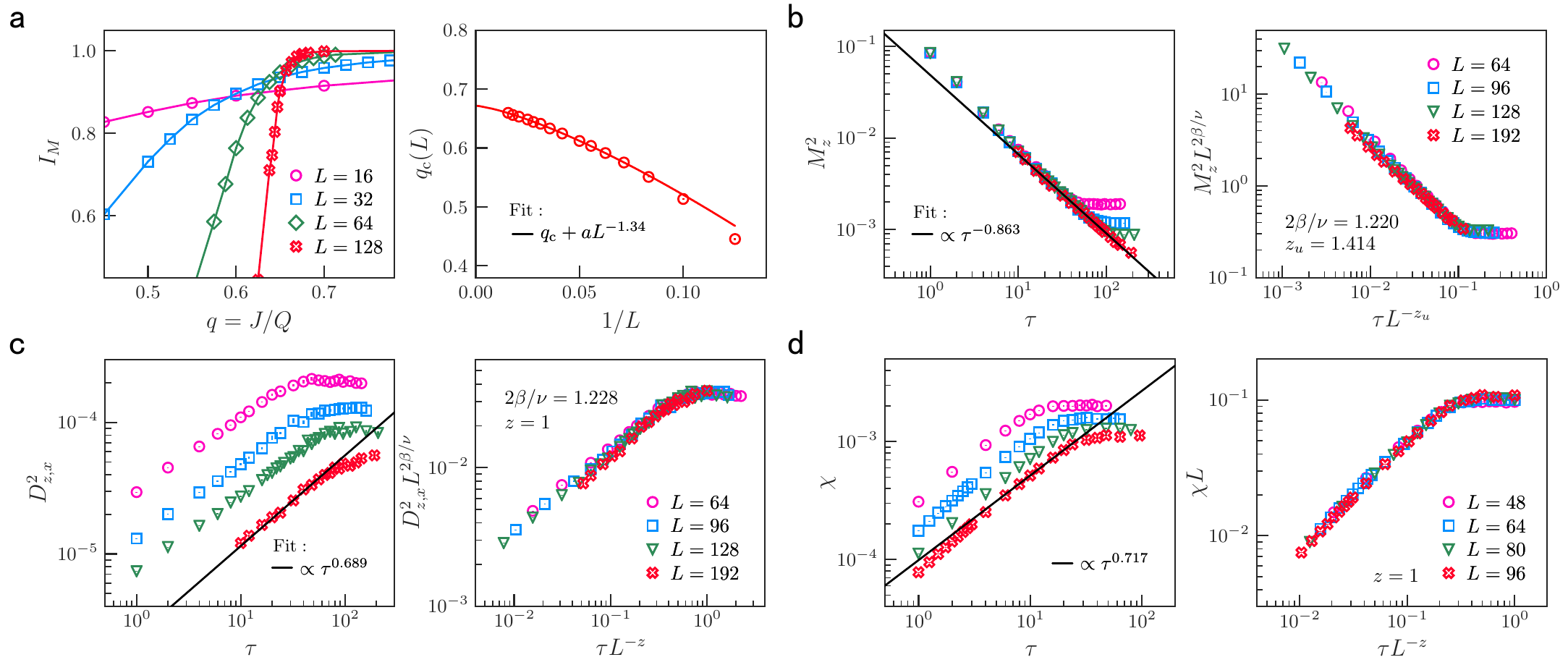}
  \vskip-3mm
  \caption{\textbf{Relaxation dynamics with the saturated N\'{e}el initial state.} \textbf{a,} The average sign of the N\'{e}el order parameter $I_M$ is used to estimate the critical point. For different $L$ and fixed $\tau L^{-1}=1/4$, crossing points (left) of curves of $I_M-q$ are extrapolated to estimate the critical point as $q_c\simeq0.671(2)$ (right), corroborating that from $I_D$ shown in Fig.~\ref{fig:vbs}. \textbf{b,} At $q=q_c$, in the long-time stage, $M^2\propto L^{-1.220}$, suggesting that $2 \beta/\nu\simeq1.220(6)$ (Supplementary Information), while in the short-time stage, $M^2\propto \tau^{-0.863}$, suggesting the exponent is $2\beta/\nu z_u$. FSS collapse (right) confirms the scaling form $M^2(\tau,L)=\tau^{-2\beta/\nu z_u}f(\tau L^{-z_u})$. Here, the subscript $z$ of $M$ means its $z$-component.  \textbf{c,} In the short-time region, $D^2$ increases as $D^2\propto \tau^{0.689}$ (left), this exponent is also close to ${(d-2\beta/\nu)/z}$ albeit with relatively large error (Supplementary Information), and $D^2\propto L^{-2}$ for fixed $\tau$ (supplemental information). The whole relaxation satisfies $D^2(\tau,L)=L^{-d}\tau^{d/z-2\beta/\nu z}f(\tau L^{-z})$, verified by FSS collapse (right). Here, the subscript $z,x$ of $D$ means its component with $z$-direction spin and $x$-direction bond. \textbf{d,} In the short-time stage, the susceptibility $\chi$ increases as $\chi\propto \tau^{\nu/\nu'}$ and depends on $L$ as $\chi\propto L^{-(1+\nu/\nu')}$  (left) (Supplementary Information). The scaling form $\chi(\tau,L)=L^{-1}f(\tau L^{-z})$ is verified by scaling collapse (right). These results show that $I_M$ and $D^2$ is controlled by the dynamics of $\xi'$, $M^2$ is controlled by the dynamics of $\xi$, and $\chi$ is hiddenly controlled by both. Note that $M^2$ and $D^2$ exchange their scaling forms compared with those from the VBS initial state as shown in Fig.~\ref{fig:vbs}.
  }
  \label{fig:antiferro}
\end{figure*}

Generally, the relaxation dynamics of the DQCP is controlled by the changes of two relevant length scales $\xi'\propto \tau^{\frac{1}{z}}$ and $\xi\propto \tau^{\frac{1}{z_u}}$. For an operator $Y$ its dynamic scaling should satisfy
\begin{equation}
\label{eq:operator}
Y(\tau,\delta,L,\{X\})=\tau^{\frac{s}{\tilde{z}}}f(\delta \tau^\frac{1}{\tilde{\nu} \tilde{z}},\tau L^{-z},\tau L^{-z_u},\{X \tau^{-\frac{c}{\tilde{z}}}\}),
\end{equation}
in which $s$ is the exponent related to $Y$, $\delta\equiv q-q_c$ with $q_c$ the critical point, $L$ is the lattice size, and $\tilde{z}$ is the dynamic exponent, which can be $z$ or $z_u$, or their combination, depending on the dynamic process, similarly, $\tilde{\nu}$ can be $\nu$ or $\nu'$, and $\{X\}$ with its exponent $c$ represents other possible relevant variables associated with the initial state. For saturated ordered and completely disordered initial states, ${X}$ vanishes since these states keep invariant under scale transformation~\cite{Yin2014,Janssen1989}. If $z_u=z$, equation~(\ref{eq:operator}) recovers the usual single-length-scale SITD scaling theory, in which, for instance, a dimensionless variable at $\delta=0$ is a function of $\tau L^{-z_u}$, and the order parameter scales as $M^2=\tau^{-\frac{2\beta}{\nu z_u}}f({\tau L^{-z_u}})$ for a saturated initial state and $M^2=L^{-d}\tau^{\frac{d}{z_u}-\frac{2\beta}{\nu z_u}}f({\tau L^{-z_u}})$ for a disordered initial state~\cite{Albano2011}.

We then explore the dynamic scaling in model~(\ref{eq:hamiltonian}) from a saturated VBS state. For a dimensionless quantity $I_D(\tau,\delta,L)$, defined as the average sign of the VBS order parameter, $I_D\equiv\langle{{{\rm sgn}\{D(\tau)\}}}\rangle$, we find in Fig.~\ref{fig:vbs}~{\bf c} that for a fixed $\tau L^{-z}$ the convergence of crossing-points of $I_D(\tau,\delta,L)$ for large $L$ corroborates the value of the critical point $q_c\simeq 0.671(2)$~\cite{Lou2009}, which indicates that $\tau L^{-z}$ dominates the dynamics of $I_D$, rather than $\tau L^{-z_u}$. Otherwise, no convergence should be seen for fixed $\tau L^{-z}$. This makes the scaling form of $I_D$ remarkably different from the usual one since the relevant length scale is not the conventional $\xi$ but the spinon confinement length $\xi'$. A possible reason is that $I_D$ is closely related to the domain walls separating regions with different VBS state, of which the thickness is characterized by $\xi'$.

\begin{figure*}[htbp]
\centering
  \includegraphics[width=\linewidth,clip]{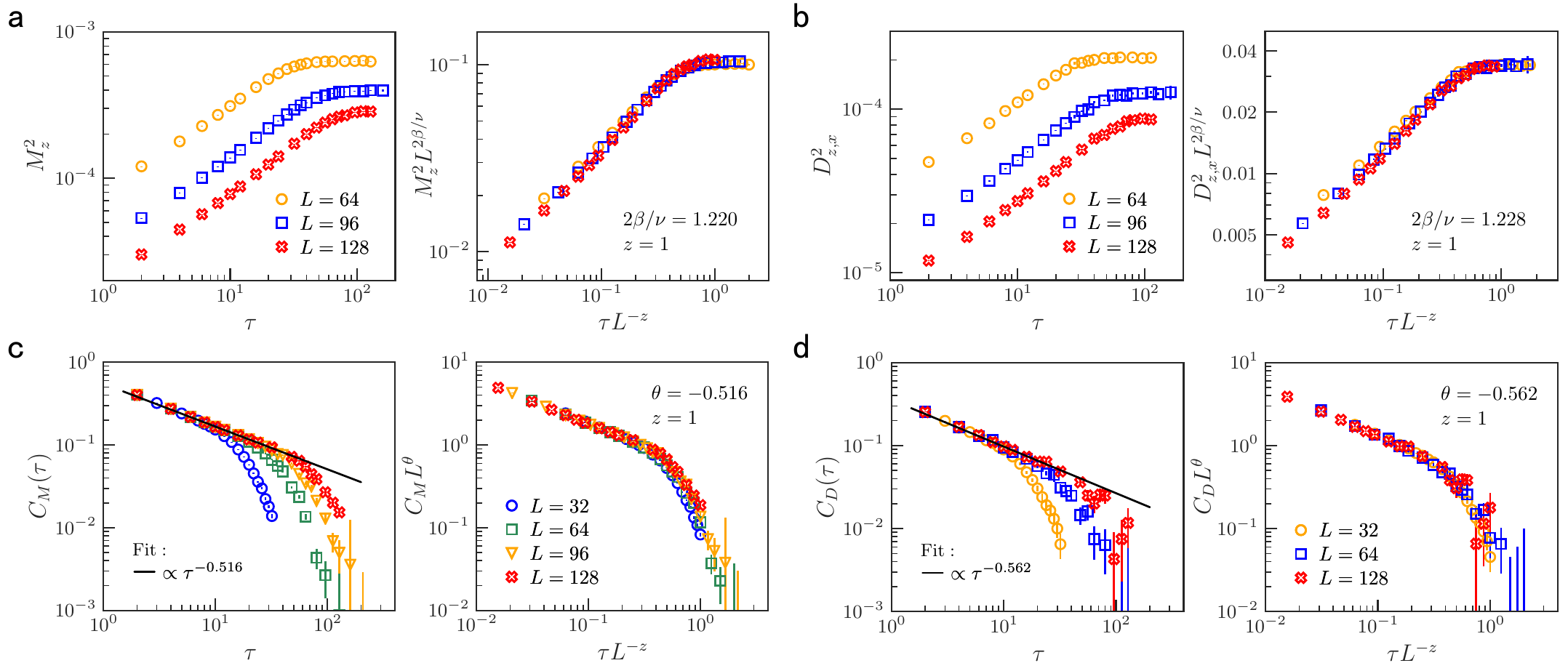}
  \vskip-3mm
  \caption{\textbf{Relaxation dynamics with the disordered initial state.} \textbf{a,b,} In the short-time stage, both $M^2$ and $D^2$ satisfy $P^2(\tau,L)=L^{-d}\tau^{d/z-2\beta/\nu z}$ in which $P$ represents $M$ or $D$ (left) and the whole relaxation dynamics satisfy the scaling form $P^2(\tau,L)=L^{-d}\tau^{d/z-2\beta/\nu z}f(\tau L^{-z})$ (right) (Supplementary Information). \textbf{c,d,} The time-correlations $C_P$ show critical initial slip behaviors, i.e., in the short-time stage, $C_P\propto \tau^\theta$ with $\theta$ being the critical initial slip exponent. For $M$, $\theta\simeq-0.516(5)$ and for $D$, $\theta\simeq-0.562(23)$. They are close to each other. The whole relaxation dynamics satisfies the scaling form $C_P=\tau^{\theta}f(\tau L^{-z})$, which are verified in the respective right panels. These results confirm the dual properties of the dynamic scaling.}
  \label{fig:initialslip}
\end{figure*}

In equilibrium, $D^2$ and $M^2$ were shown to satisfy similar scaling forms~\cite{Lou2009,Harada2013,Nahum2015b}. Strikingly, here we find that their relaxation dynamics from a saturated VBS initial state are controlled by different length scales. For $D^2$, Fig.~\ref{fig:vbs}~{\bf d} shows that it obeys $D^2(\tau,L)=\tau^{-\frac{2\beta}{\nu z_u}}f(\tau L^{-z_u})$ for which $\xi\propto \tau^{-\frac{1}{z_u}}$ dominates, similar to the usual single-length-scale case. In the short-time stage, $D^2(\tau,L)\propto \tau^{-\frac{2\beta}{\nu z_u}}$ and it is almost independent of $L$, while its long-time limit recovers the equilibrium scaling $D^2\propto L^{-\frac{2\beta}{\nu}}$. The scaling form of $D^2$ meets these two conditions simultaneously provided that $f(\tau L^{-z_u})\sim (\tau L^{-z_u})^{\frac{2\beta}{\nu z_u}}$ for $\tau\rightarrow\infty$. Otherwise, if $\xi'$ dominates, the appearance of $\tau L^{-z}$ will make it difficult for the scaling function to satisfy these two limits simultaneously with a simple form. The above scaling form is confirmed by rescaling collapse in Fig.~\ref{fig:vbs}~{\bf d}. Contrarily, for $M^2$, Fig.~\ref{fig:vbs}~{\bf e} shows that its short-time dynamics obeys $M^2\propto L^{-d}\tau^{\frac{d}{z}-\frac{2\beta}{\nu z}}$ (Supplementary Information), requiring its full scaling form to be $M^2(\tau,L)=L^{-d}\tau^{\frac{d}{z}-\frac{2\beta}{\nu z}}f(\tau L^{-z})$, where $\xi'$ and $\tau L^{-z}$ dominate. A possible reason for this discrepancy is that $D^2$ is deeply affected by the memory from the initial state, whereas $M^2$ only feels a disordered initial state (Supplementary Information).

More interestingly, some quantities can even show fascinating relaxation behaviors controlled by the dynamics of $\xi$ and $\xi'$ simultaneously. For instance, in equilibrium, the VBS domain wall energy density $\kappa$ is found to scale as $\kappa\propto \xi'^{-1}\xi^{-(d+z-2)}$ ($d+z-2=1$)~\cite{Senthil2004,Shao2016}. Here we generalized this scaling into the nonequilibrium case. As shown in Fig.~\ref{fig:vbs}~{\bf f}, $\kappa$ relaxes according to $\kappa(\tau,L)=\tau^{-\frac{1}{z}}\tau^{-\frac{1}{z_u}}f(\tau L^{-z})$ from the satuarated VBS state at $q=q_c$. In the short-time stage, $\kappa\propto \tau^{-\frac{1}{z}}\tau^{-\frac{1}{z_u}}$, while in the long-time stage, $\kappa$ crosses over to $\kappa\sim L^{-(1+\frac{\nu}{\nu'})}$ as $f(\tau L^{-z})\sim (\tau L^{-z})^{\frac{1}{z_u}+\frac{1}{z}}$ for $\tau\rightarrow\infty$, recovering its equilibrium finite-size scaling~\cite{Senthil2004b,Shao2016,Shao2015} (Supplementary Information).

To illustrate the role played by the initial state in the relaxation dynamics, we now study the dynamic scaling with the saturated antiferromagnetic initial state. From Fig.~\ref{fig:antiferro}~{\bf a}, we find that the average sign of the N\'{e}el order parameter defined as $I_M\equiv\langle{{{\rm sgn}\{M(\tau)\}}}\rangle$ obeys $I_M=f(\tau L^{-z})$ similar to $I_D$. The critical point estimated by crossing-point analyses of $I_M$ is $0.671(2)$, consistent with that given by $I_D$. Notably, we find that $M^2$ and $D^2$ exchange their scaling forms compared with the VBS initial state case. Namely, $M^2$ satisfies $M^2(\tau,L)=\tau^{-\frac{2\beta}{\nu z_u}}f(\tau L^{-z_u})$ governed by $\xi$, and $D^2$ satisfies $D^2(\tau,L)=L^{-d}\tau^{\frac{d}{z}-\frac{2\beta}{\nu z}}f(\tau L^{-z})$ governed by $\xi'$, as shown in Fig.~\ref{fig:antiferro}~{\bf b,c}, respectively. In addition, we find in Fig.~\ref{fig:antiferro}~{\bf e} that the dynamics of the susceptibility $\chi$ at the momentum $(\frac{2\pi}{L},0)$~\cite{Shao2016} satisfies $\chi(\tau,L)=L^{-(d-z)}f(\tau L^{-z})$ (here $d-z=1$). An interesting phenomenon is that in the short-time stage $\chi(\tau,L)\sim L^{-1}(\tau L^{-z})^\frac{\nu}{\nu'z}$, indicating a hidden interplay between the dynamics of two length scales (Supplmentary Information).

The exchange of scaling forms for $M^2$ and $D^2$ exhibits an enchanting dual dynamic scaling behavior, which reflects the imprint of emergent symmetry in nonequilibrium dynamics. Its appearance can be traced back to the intertwining relation between the N\'{e}el order and the VBS order, i.e., the N\'{e}el order can be regarded as the condensation of the spinon which stays at the intersection of the VBS domain walls; while the VBS order can be regarded as the condensation of the hedgehog topological defects of the N\'{e}el phase~\cite{Senthil2004,Senthil2004b}. In this context, $\xi'$ is not only the typical distance of spinons which make up the N\'{e}el order parameter from the VBS background, but also measures the thickness of the VBS domain walls which are constructed by the quadrupled monopole events in the antiferromagnetic background and characterizes the VBS fluctuations. This may explain why $M^2$ ($D^2$) is controlled by the dynamics of $\xi'$ when the initial state is the saturated VBS (N\'{e}el) state.

The dual dynamic scaling also manifests itself when the initial state is a disordered state. We find in Fig.~\ref{fig:initialslip}~{\bf a,b} that $M^2$ and $D^2$ evolves according to the same scaling form $P^2(\tau,L)=L^{-d}\tau^{\frac{d}{z}-\frac{2\beta}{\nu z}}f(\tau L^{-z})$ in which $P$ represents $M$ or $D$ and $\xi'$ dominates their relaxation dynamics. Moreover, for this initial state, the SITD features another characteristic phenomenon dubbed the critical initial slip (CIS)~\cite{Janssen1989,Yin2014}. The CIS in the usual LGW criticality shows that $C_P(\tau)\equiv L^{d}\langle{P(0)P(\tau)\rangle}$ obeys $C_P(\tau,L)=\tau^{\theta}f(\tau L^{-z_u})$ with $\theta>0$~\cite{Tome1998}. In contrast, in the DQCP, we find in Fig.~\ref{fig:initialslip}~{\bf c,d} that $\xi'$ governs the CIS behavior and $C_P$ obeys $C_P(\tau,L)=\tau^{\theta}f_{C}(\tau L^{-z})$ in which both $\theta$ are negative, i.e., $\theta\simeq -0.516(5)$ for $M$ and $\theta\simeq -0.562(23)$ for $D$ (Supplementary Information). In the LGW criticality, the reason for a positive $\theta$ is that the real critical point usually shifts towards the ordered phase compared with its mean-field value due to critical fluctuations~\cite{Janssen1989,Yin2014}. However, in the DQCP, both phases are ordered phases and the fluctuations are well-matched in both sides of the critical point due to the emergent symmetry. This explains the negative $\theta$. In addition, the value of $\theta$ for $M$ is close to that for $D$, demonstrating again the dual dynamic scaling.

To conclude, our work studies the nonequilibrium dynamics in $2$D DQCP and finds that the interplay between the deconfined process and the fluctuating modes can contribute striking nonequilibrium properties. Our present results could be realized in noisy-intermediate-scale quantum computers (e.g. Regetti), for which the imaginary-time evolution algorithm was proposed recently~\cite{Motta2020}. Our work also provides significant instructions in the real-time dynamics of the DQCP by noting that the short-time scaling also manifests itself in real-time dynamics in various quantum systems~\cite{Cardy2006}. Moreover, our results can be generalized to other systems featuring exotic deconfinement dynamics and boundary criticality of DQCP.

\clearpage
\noindent {\bf Methods}

The imaginary-time relaxation results for the J-Q$_{3}$ model were obtained using the projector QMC method. The method is well documented in the literature~\cite{Sandvikrev}. Here we only give brief overviews, particularly focusing on its application in the imaginary-time dynamics~\cite{Liu2013,Farhi2012}.

For a given starting state $|\psi(0)\rangle$ at $\tau=0$, the imaginary-time evolution of the state obeys the Schr{\"o}rdinger dynamics which gives $|\psi(\tau)\rangle=e^{-\tau H}|\psi(0)\rangle$.
In the projector method, the imaginary-time evolution operator $U(\tau)=e^{-\tau H}$ is Taylor-expanded into powers of the Hamiltonian $H^{n}$.
The normalization $Z=\langle\psi(0)|e^{-2\tau H}|\psi(0)\rangle$ is importance sampled using the overlap of the bra and ket states as the sampling weight in the standard $S^{z}$ basis or other basis such as the nonorthogonal valence bond (VB) basis.
The expansion power $n$ is updated and automatically truncated to some maximum length without detectable errors.
A full Monte Carlo sweep consists of local diagonal updates and global off-diagonal updates. The local diagonal updates replace unit operators with diagonal ones with appropriate acceptance rate and vice versa in the operator sequence. The global operator-loop updates used here switch the operator types from diagonal to off-diagonal or vice versa and update the corresponding states along the propagation direction with probability $1/2$. Detailed balance and ergodicity are mantained. The computational consumption of a full sweep of Monte Carlo update scales as $2\tau N$ with $N=L^{d}$.
Expectation value for an operator $O$ at imaginary time $\tau$ is given by $\langle O(\tau)\rangle = \langle\psi(\tau)|O|\psi(\tau)\rangle/Z$, where $\langle\cdots\rangle$ represents nonequilibrium average.

In the projector QMC method, the intial states are realized by controlling the boundaries of the projection direction, for instance, to enforce antiferromagnetic order in the initial state, the loops touching the imaginary-time boundaries in the operator-loop updates should be fixed. Regarding the initial state required, different bases are applied. For the VBS initial state, the VB basis is used as it restricts the sampling in the singlet sector. In this basis, the initial state is expressed in terms of valence bonds, which makes it advantageous for VBS initial states. For the antiferromagnetic or disordered initial state, the simulations are carried out standard $S^{z}$ basis.

To study the relaxation dynamics, we investigate the imaginary-time evolution of different physical quantities under different initial conditions, including the N{\'e}el and VBS order parameters, domain wall energy density $\kappa$, spinon size $\Lambda$, susceptibility $\chi$.

The N{\'e}el and VBS order parameters are defined as ${\bf M}=\sum_{r}(-1)^{r}{\bf S}_{r}/N$ and $D_{x}=\sum_{r}(-1)^{r_{x}}({\bf S}_{r}\cdot{\bf S}_{r+\hat{x}})/N$, respectively. The subscript $x$ in $D_{x}$ represents $x$-oriented bond order and $\hat{x}$ is the unit lattice vector in the $x$ direction. The squared order parameters $M^{2}$ and $D_{x}^{2}$ are defined accordingly.
In the VB basis, improved estimators for $M^{2}$ and $D^{2}_{x}$ are avaliable, which are expressed in the context of loops formed by valence bonds.
In the $S^{z}$ basis, we compute the squared $z$-component of the VBS order parameter $D^{2}_{z,x}$
which has the same scaling form of the full VBS order parameter. To ensure the convergence, we analyze the imaginary-time dependence of the order parameters until the results only fluctuate within statistical errors.


For the density of domain wall energy $\kappa$, different boundary conditions are employed in order to introduce twisted boundaries~\cite{Shao2016method}. In the background of $y$-oriented dimers, shifting one of the two $x$-edges by one lattice spacing brings a twist angle of $\Delta\phi=\pi$. The periodic boundary condition in the $x$ direction is broken. The energy $E(\Delta\phi)$ difference between the system with and without twisted boundaries defines the size-normalized domain wall energy density, namely, $\kappa=(E(\pi)-E(0))/L$.
Analogously, rotating one of the two $x$-edges from $y-$ into $x$-oriented bond order causes $\Delta\phi=\pi/2$ (the boundaries are illustrated in Supplementary Information). In the projector QMC method, $E$ is simply given by the average number of non-unit operators, namely $E=-\langle n\rangle/2\tau$.

The computions of the spinon size $\Lambda$ are done in the $S=1$ sector with two unpaired spins in the VBS background~\cite{Shao2016method}.
The unpaired spin in the bra and ket states is connected via a string, which represents a spinon. The two spinon distance is found capable in capturing the faster divergence of $\xi^\prime$. The definition of $\Lambda$ is associated with the lattice sites covered by the string. For the two spinons covering $n_{1}$ and $n_{2}$ sites, we compute $\langle r^{2}\rangle=\sum_{i}^{n1}\sum_{j}^{n2}r_{ij}^{2}/n_{1}n_{2}$ with $r_{ij}^{2}=|\vec{r}_{1}(i)-\vec{r}_{2}(j)|^{2}$ being the squared distance of the two points $i$ and $j$ located on the two strings. The spinon size $\Lambda$ is then defined as $\Lambda=\langle r^{2}\rangle^{1/2}$.

The susceptibility $\chi$ studied here is defined via the fourier transformation of the local correlations $G(i,j)=\partial{\langle S_{i}^{z}\rangle}/\partial{\Gamma_{j}}$ at zero external field $\Gamma_{j}=0$, which measures the response of spin $i$ to the external field $\Gamma_{j}$ at site $j$.
In the projector QMC method, $\chi({\bf q})$ is given by~\cite{Sandvikrev}
\begin{eqnarray*}
  \chi({\bf q})=\frac{\tau}{N}\langle\sum_{m=0}^{n}\frac{w(n,m)}{n+1-m}M_{\bf q}(m)\sum_{p=0}^{n}M_{-{\bf q}}(p)\\
  +\sum_{m=0}^{n}\frac{w(n,m)}{n+1-m}M_{\bf q}(m)M_{-{\bf q}}(m)\rangle,
\end{eqnarray*}
in which $M_{\bf q}=\sum_{i}S_{i}^{z}e^{-i{\bf q}\cdot{\bf r}_i}$ and $w(n,m)=\binom{n}{m}/2^{n}$. Only the real parts should be calculated since $\chi({\bf q})$ must be real-valued.
Here we compute $\chi$ at the smallest nonzero momentum ${\bf q}=(\frac{2\pi}{L},0)$ since the uniform susceptibility $\chi({\bf q}=0)$ vanishes in the $S=0$ sector.

\noindent {\bf Acknowledgements}
We gratefully acknowledge helpful discussions with A. W. Sandvik, H. Shao, P. Ye and F. Zhong. Y.R.S. acknowledges support from Grant No. NSFC-11947035 and the startup grant (RP2020120) at Guangzhou University. S.K.J. is supported by the Simons Foundation via the It From Qubit Collaboration.  S.Y. is supported by the National Natural Science Foundation of China (Grant No. 41030090). The numerical analysis was in part performed on TianHe-2 (the National Supercomputer Center in Guangzhou).

\begin{widetext}
\clearpage

\section{Supplementary Information: Additional Numerical Results}
\renewcommand{\theequation}{S\arabic{equation}}
\setcounter{equation}{0}
\renewcommand{\thefigure}{S\arabic{figure}}
\setcounter{figure}{0}
\renewcommand{\thetable}{S\arabic{table}}
\setcounter{table}{0}

\subsection{The road to emergent symmetry}
The DQCP theory shows that near the critical point of the J-Q$_3$ model, the deconfined spinons coupled to an emergent $U(1)$ gauge field dominate the critical behaviors and the N\'{e}el order parameter ${\bf M}\equiv(M_x,M_y,M_z)$ and VBS order parameter ${\bf D}\equiv(D_x,D_y)$ should be understood as composites of these objects. Although these two order parameters are utterly inequivalent microscopically, an emergent SO$(5)$ symmetry between them arises in the infrared scale to take into account the rotation in the superspin $\mathcal{S}=(M_x,M_y,M_z,D_x,D_y)$~\cite{Senthil2004sup,Senthil2004bsup,Levin2004sup,Nahum2015bsup}. This emergent symmetry only appears at the critical point, while in the VBS phase, the system breaks this continuous symmetry into a discrete $Z_4$ symmetry~\cite{Senthil2004bsup}. Accordingly, besides the usual correlation length $\xi$, there exists another length scale $\xi'$ characterizing the spinon confinement length or the thickness of the VBS domain walls. Scaling theories based on these two length scales were proposed to explain the anomalous scaling properties near the DQCP.
\begin{figure}[htbp]
  \includegraphics[width=0.9\columnwidth,clip]{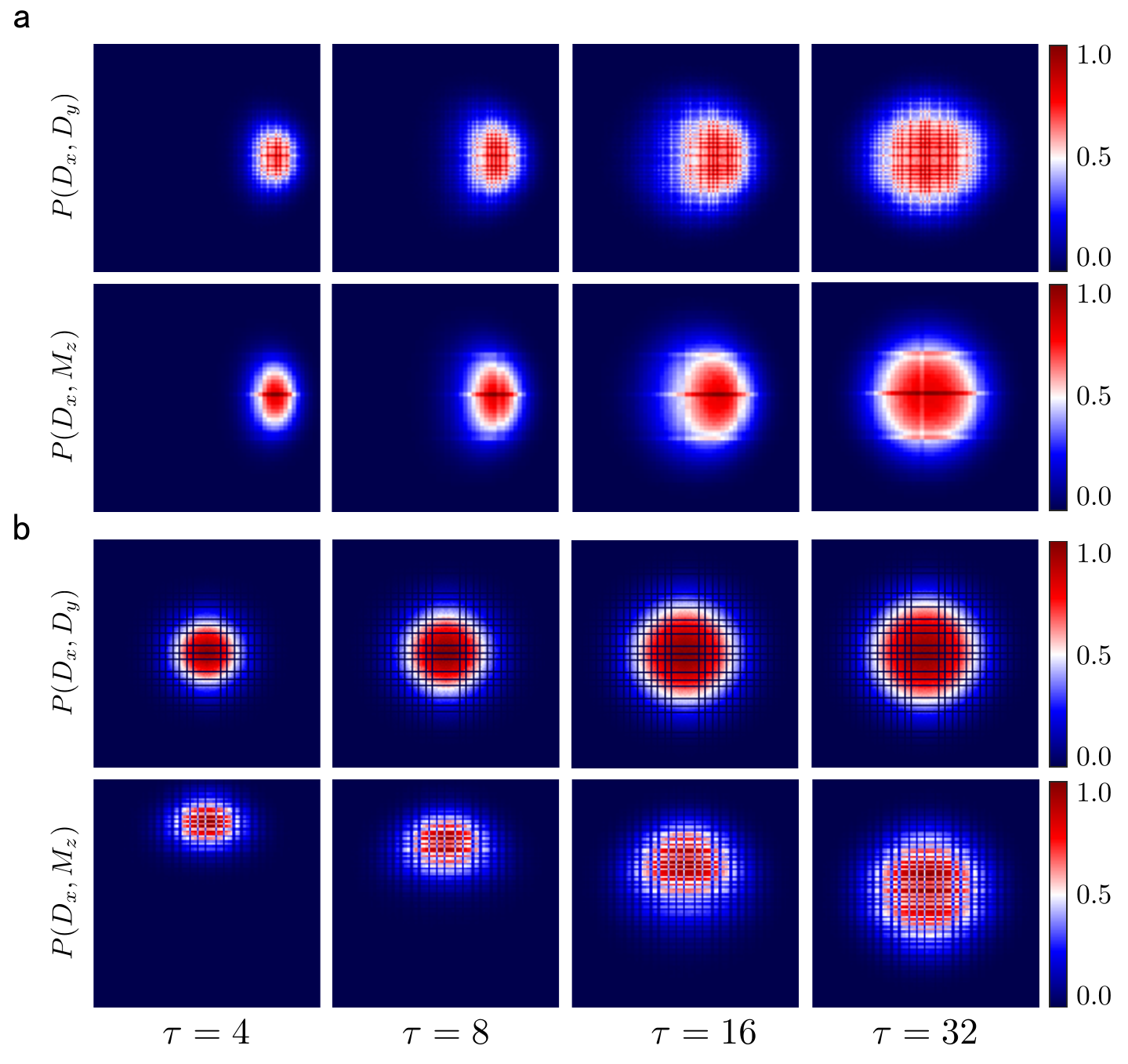}
  \vskip-3mm
  \caption{\textbf{The road to the emergent symmetry.} \textbf{a,} With a saturated initial VBS state, the histogram of the probability distribution of the order parameters shows that the SO$(5)$ symmetry emerges gradually as $\tau$ increases at the critical point (after a trivial rescaling of VBS order parameter). \textbf{b,} Similarly, with a saturated initial N\'{e}el state, the histogram of the probability distribution of the order parameters also shows that the SO$(5)$ symmetry emerges gradually as $\tau$ increases at the critical point. }
  \label{fig:esymmetry}
\end{figure}

In the imaginary-time relaxation, the energy scale flows from the ultraviolate scale down to the infrared scale. Accordingly, the generation process of the emergent symmetry can be observed. From a saturated VBS order with vanishing correlation length, we show in Fig.~\ref{fig:esymmetry}~\textbf{a} the dynamics of the appearance of the SO$(5)$ symmetry at the critical point of model~(1) by calculating numerically the evolution of the probability distribution of the order parameters. As displayed in Fig.~\ref{fig:esymmetry}~\textbf{a}, in the infancy stage, the microscopic dynamics dominates and the system breaks the Z$_4$ symmetry. Then the probability distribution of the superspin begins to spread. In the $D_x-D_y$ plane, the symmetry under continuous U$(1)$ of $\bf D$ emerges gradually, although the microscopic lattice symmetry only allows $\pi/2$ rotation. Moreover, in the $D_x-M_z$ plane, another U$(1)$ symmetry mixing the $M_z$ and $D_x$ emerges gradually from an apparent symmetry-breaking initial state. By taking into account the SO$(3)$ symmetry of $\bf M$, we demonstrate the relaxation dynamics to the emergent SO$(5)$ symmetry. Similarly, Fig.~\ref{fig:esymmetry} \textbf{b} shows the generation process of the emergent SO$(5)$ symmetry with a saturated N\'{e}el initial state.

\subsection{Scaling properties of the N\'{e}el and VBS order parameters}
First, we show the scaling properties of $M^2$ and $D^2$ in the long-time equilibrium limit $\tau\rightarrow\infty$. In equilibrium, it was shown that at the critical point $M^2$ satisfies $M^2\propto L^{-2 \beta/\nu}$ and $D^2$ also satisfies $D^2\propto L^{-2 \beta/\nu}$ with almost the same exponents. In Fig.~\ref{fig:equiMD}, we verify these scaling properties.

Second, we explore the dependence of the dynamics of $M^2$ on the lattice size $L$ with the initial state being the saturated VBS state. In Fig.~\ref{fig:mvbssize}, we find that in the short-time relaxation stage $M^2\propto L^{-2}$ for a fixed $\tau$, confirming $M^2\propto L^{-d}$. Combining the dependence of $M^2$ on $\tau$, we confirm that in the short-time stage, $M^2$ satisfies $M^2\propto L^{-d}\tau^{\frac{d}{z}-\frac{2\beta}{\nu z}}$. Similarly, from a saturated N\'{e}el state, the VBS order parameter satisfies $D^2\propto L^{-d}$ for fixed $\tau$ in the short-time stage as shown in Fig.~\ref{fig:vbsmsize}. Accordingly, in the short-time region, $D^2$ obeys $D^2\propto L^{-d}\tau^{\frac{d}{z}-\frac{2\beta}{\nu z}}$. In addition, when the initial state is the disordered state, both $M^2$ and $D^2$ are proportional to $L^{-2}$ for a fixed time in the short-time stage as shown in Fig.~\ref{fig:mvbsdisorder}, demonstrating their short-time relaxation dynamics satisfy $M^2\propto D^2\propto L^{-d}\tau^{\frac{d}{z}-\frac{2\beta}{\nu z}}$.

Third, as we discussed above, the relaxation dynamics of $M^2$ from the saturated VBS initial state satisfies $M^2\propto L^{-d}\tau^{\frac{d}{z}-\frac{2\beta}{\nu z}}$ in the short-time stage. We find this scaling behavior is well captured for large system sizes. However, for small systems, finite size corrections should be included. We extract the exponent of $\tau$ for different system sizes as shown in Fig.~\ref{fig:mvbsslope}. It is expected this exponent will be saturated at $\frac{d}{z}-\frac{2\beta}{\nu z}$ for large $L$. Similar behavior of the exponent also happens for the dynamics of $D^2$ from the saturated N\'{e}el initial state, as shown in Fig.~\ref{fig:mvbsslope}.
\begin{figure}[htbp]
  \includegraphics[width=0.5\columnwidth,clip]{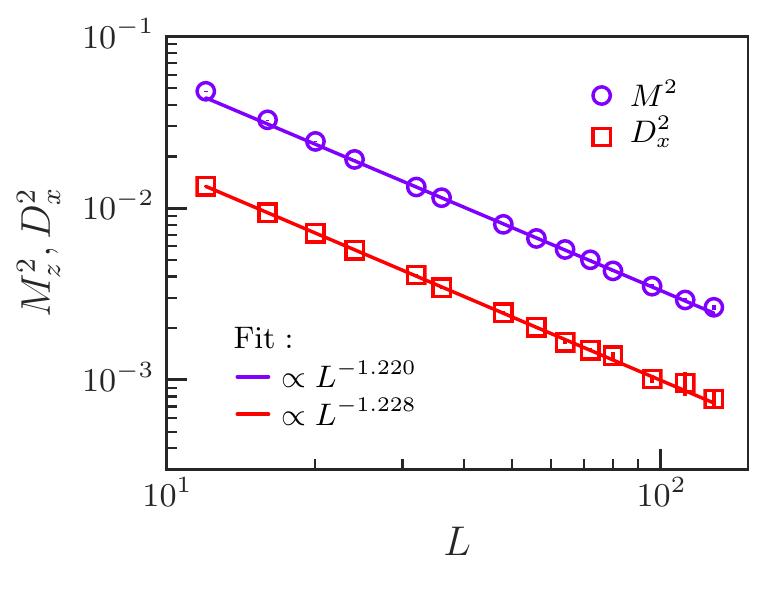}
  \vskip-3mm
  \caption{\textbf{Equilibrium finite-size scaling of $M_z^2$ and $D_x^2$ at the critical point}. For large $\tau$, the system tends to the equilibrium state and the order parameters becomes their equilibrium values. They are independent of $\tau$ and only depend on $L$. Power fitting shows that $M^2\propto L^{-1.228}$ and $D^2\propto L^{-1.220}$. These two exponents are just $2\beta/\nu$. They are close to each other, indicating the appearance of the emergent symmetry.}
  \label{fig:equiMD}
\end{figure}

\begin{figure}[htbp]
  \includegraphics[width=0.5\columnwidth,clip]{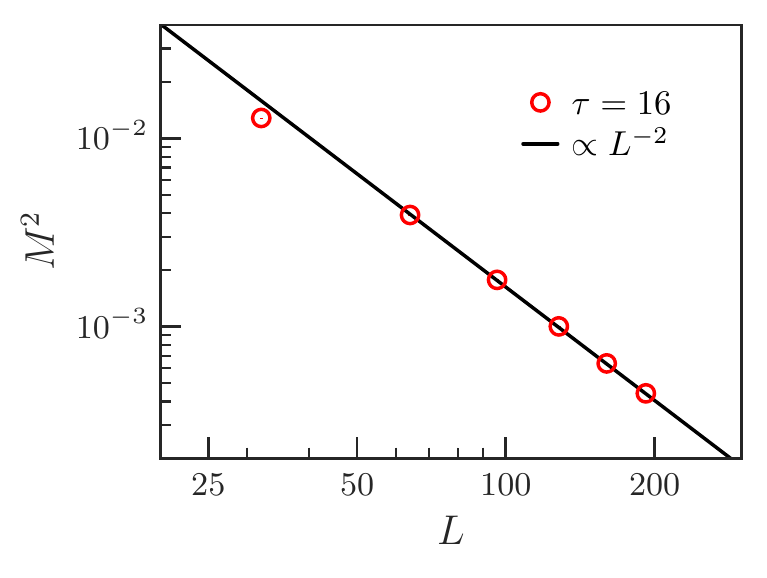}
  \vskip-3mm
  \caption{\textbf{Dependence of $M^2$ on $L$ for fixed $\tau$ in the short-time relaxation stage with the saturated VBS initial state.} For different lattice sizes $L$ and $\tau=16$, $M^2\propto L^{-2}$. The exponent $2$ is just the spatial dimension of the system.}
  \label{fig:mvbssize}
\end{figure}

\begin{figure}[htbp]
  \includegraphics[width=0.5\columnwidth,clip]{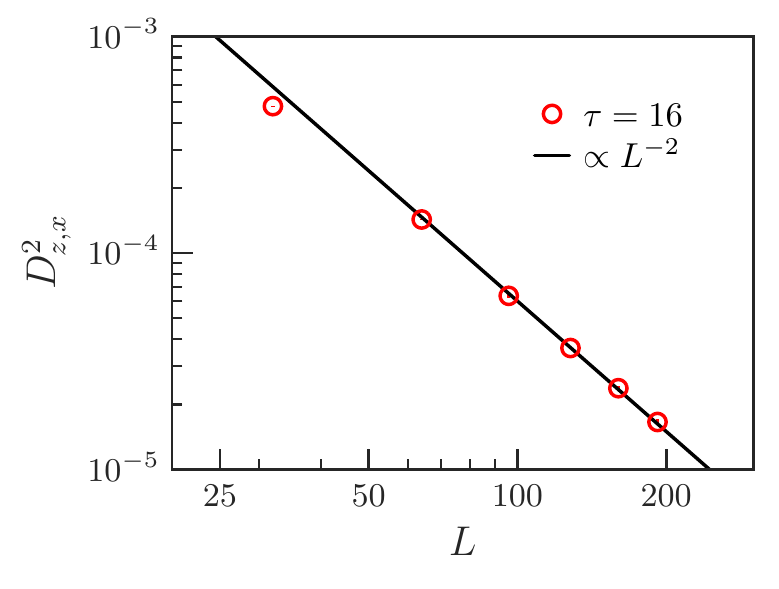}
  \vskip-3mm
  \caption{\textbf{Dependence of $D^2$ on $L$ for fixed $\tau$ in the short-time relaxation stage with the saturated N\'{e}el initial state.} For different lattice sizes $L$ and $\tau=16$, $D^2\propto L^{-2}$. The exponent $2$ comes from the spatial dimension of the system. }
  \label{fig:vbsmsize}
\end{figure}

\begin{figure}[htbp]
  \includegraphics[width=0.5\columnwidth,clip]{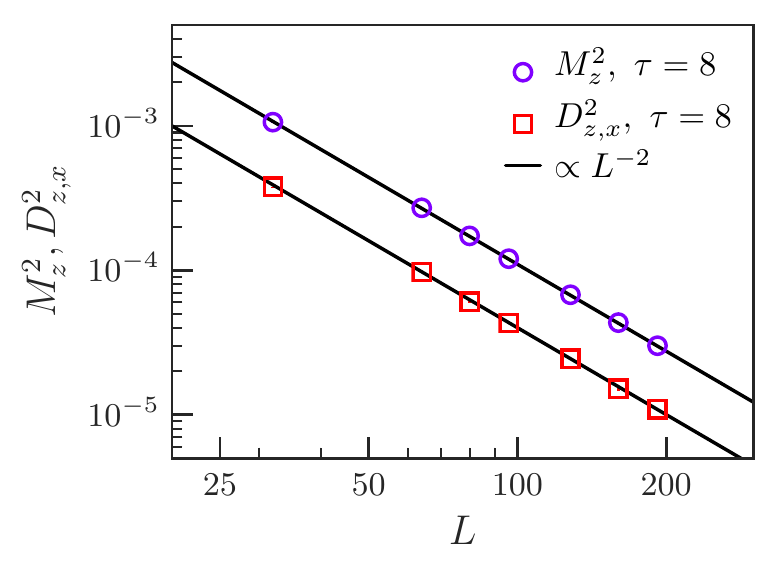}
  \vskip-3mm
  \caption{\textbf{Dependence of $M^2$ and $D^2$ on $L$ for fixed $\tau$ in the short-time relaxation stage with the completely disordered initial state.} For different lattice sizes $L$ and $\tau=8$, $D^2\propto M^2 \propto L^{-2}$. The exponent $2$ comes from the spatial dimension of the system. }
  \label{fig:mvbsdisorder}
\end{figure}

\begin{figure}[htbp]
  \includegraphics[width=0.5\columnwidth,clip]{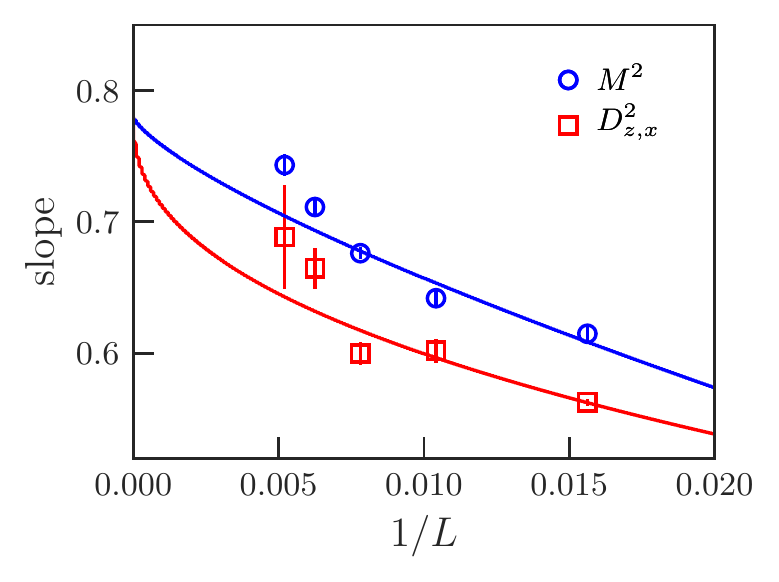}
  \vskip-3mm
  \caption{\textbf{Dependence of the exponent of $\tau$ in $M^2$ ($D^2$) on $L$ in the short-time relaxation stage with the VBS (N\'{e}el) initial state.}}
  \label{fig:mvbsslope}
\end{figure}

In above cases, by assuming that the short-time dynamics and the long-time equilibrium dynamics can be described by a unified scaling form, we find that the order parameter $M^2$ in the whole relaxation process obeys $M^2=L^{-d}\tau^{\frac{d}{z}-\frac{2\beta}{\nu z}}f(\tau L^{-z})$, and $D^2$ satisfies the similar scaling form only with a different scaling function. In these scaling forms, the scaling function $f(\tau L^{-z})\sim const$ with $const$ being a constant when $\tau\ll L^{z}$, while $f(\tau L^{-z})\sim (\tau L^{-z})^{\frac{2 \beta}{\nu z}-\frac{d}{z}}$ when $\tau\gg L^{z}$. Although this scaling form is similar in form to the conventional corresponding scaling theory~\cite{Albano2011sup}, they are remarkably different from each other, since in the present case, $z$ represents the scaling relation between the time and the spinon confinement length instead of the usual correlation length.

Here, we want to emphasize that although for the present range of lattice sizes the scaling form works quite well, more complicated scaling behaviors can not ruled out for large system size. Therein both $\tau L^{-z}$ and $\tau L^{-z_u}$ are possibly involved simultaneously. However, to probe these possible complex scaling behaviors, more powerful computational resources are needed. And we leave them for further studies.

\subsection{Scaling properties of the density of the domain wall energy $\kappa$}
The density of domain wall energy $\kappa$ can be calculated by the energy difference between ground states with and without domain walls~\cite{Shao2016sup}. Figure~\ref{fig:boundarycon} illustrates two different boundary conditions of the VBS domain walls with two different twist angles. In the left panel of Fig.~\ref{fig:boundarycon}, the VBS order has a relative shift of one lattice spacing between two opposite edges. This corresponds to the twist angle $\Delta\phi=\pi$ of the VBS order parameter. In the right panel of Fig.~\ref{fig:boundarycon}, the VBS order has vertical and horizontal dimers on two opposite sides. This corresponds to the twist angle $\Delta\phi=\pi/2$ of the VBS order parameter.

\begin{figure}[htbp]
  \includegraphics[width=0.5\columnwidth,clip]{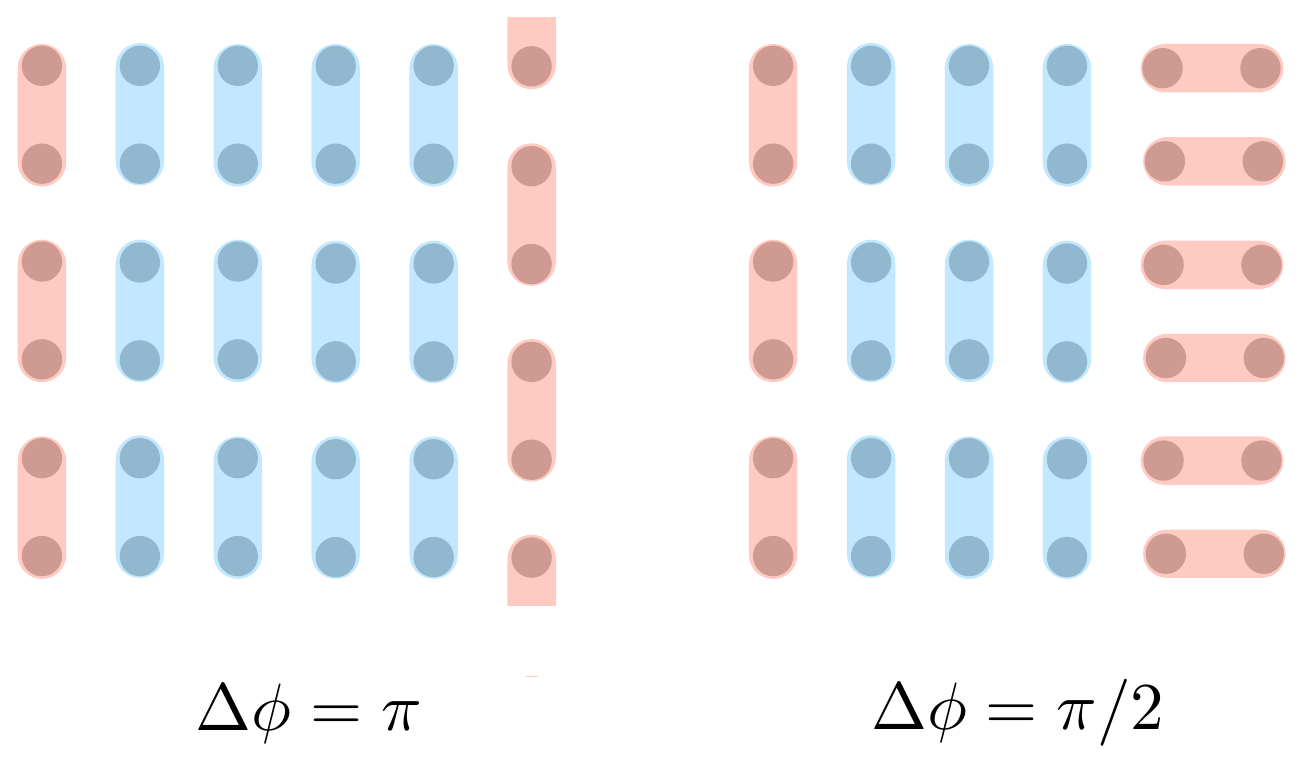}
  \vskip-3mm
  \caption{\textbf{Boundary conditions for the calculation of the domain wall energy density $\kappa$}. Boundary states with a twist angle $\Delta\phi=\pi$ and $\Delta\phi=\pi/2$ are shown in the left and right panels, respectively. The blue bonds represent the bulk state and the red bonds represent the edge state. The open boundary condition is chosen in the horizontal direction and the periodic boundary condition is imposed in the vertical direction.}
  \label{fig:boundarycon}
\end{figure}

In the main text, we focus on the case for $\Delta\phi=\pi$. We find that in the short-time stage, $\kappa$ changes with $\tau$ as $\kappa\propto \tau^{-\frac{1}{z}}\tau^{-\frac{1}{z_u}}$ at the critical point and its dependence on $L$  is very weak. This result is consistent with the dynamic generalization of the scaling relation $\kappa\propto \xi^{-1}\xi'^{-1}$, given that $\xi(\tau)\sim \tau^{-\frac{1}{z_u}}$ and $\xi'(\tau)\sim \tau^{-\frac{1}{z}}$. In the long-time equilibrium limit, $\kappa$ recovers its equilibrium scaling $\kappa\propto L^{-(1+\frac{\nu}{\nu'})}$. These two limits impose that in the whole process, $\kappa$ should satisfy $\kappa(\tau,L)=\tau^{-\frac{1}{z}}\tau^{-\frac{1}{z_u}}f(\tau L^{-z})$. In the main text, this scaling form is confirmed by scaling collapse for different system sizes.

\begin{figure}[htbp]
 \includegraphics[width=0.5\columnwidth,clip]{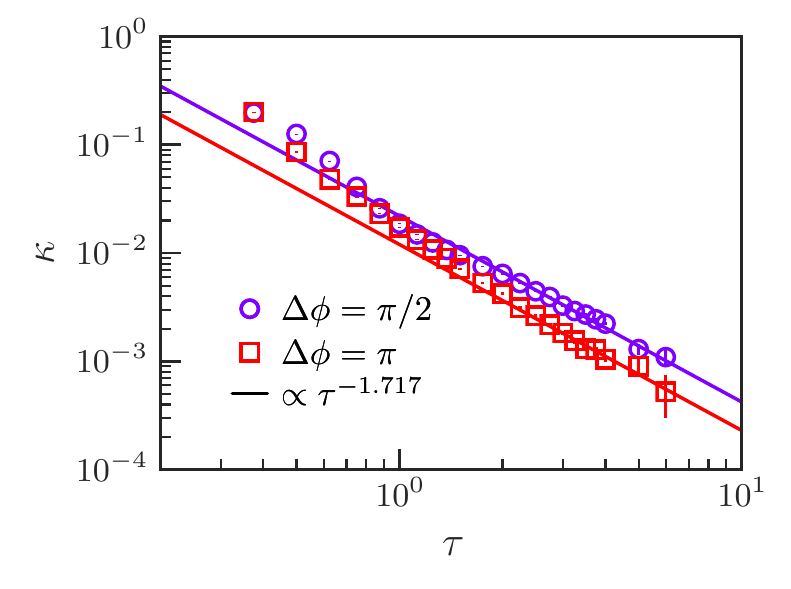}
  \vskip-3mm
  \caption{\textbf{Dependence of $\kappa$ on $\tau$ for fixed $\tau L^{-z}$ with different boundary conditions.} For $\tau L^{-z}=1/4$, $\kappa$ decays with $\tau$ as $\kappa\propto \tau^{-1.717}$ with the exponent being $1+\nu/\nu'$. This relation applies for different boundary conditions, demonstrating the universal properties.}
  \label{fig:kappafixtall}
\end{figure}
Here we check the universal properties of the scaling behaviors of $\kappa$ with a different twist angle $\pi/2$. In this case, we find that for fixed $\tau L^{-z}$ with $z=1$, $\kappa$ kappa decays with $\tau$ according to $\kappa\propto \tau^{-\frac{1}{z}}\tau^{-\frac{1}{z_u}}$, as shown in Fig.~\ref{fig:kappafixtall}. For comparison, we also plot the results for $\Delta\phi=\pi$. These results show that $\kappa$ satisfies $\kappa(\tau,L)=\tau^{-\frac{1}{z}}\tau^{-\frac{1}{z_u}}f(\tau L^{-z})$ regardless of the boundary condition, confirming its universality.

\subsection{Scaling properties of the susceptibility $\chi$}
In the main text, we show that $\chi$ increases with $\tau$ as $\chi\propto \tau^{\frac{1}{z_u}}$ for a saturated N\'{e}el initial state. Here, we show that in the short-time stage, $\chi\propto L^{-(d-z+\frac{\nu}{\nu'})}$ (here $d-z=1$) for a fixed $\tau$, as illustrated in Fig.~\ref{fig:chil}. Thus, in the short-time stage, $\chi\propto L^{-(1+\frac{\nu}{\nu'})} \tau^{\frac{z}{z_u}}$. In the long-time stage, $\chi$ satisfies $\chi\propto L^{-1}$ as shown in the main text.  Also, in the main text, the scaling collapse shows that $\chi$ in the whole relaxation process satisfies $\chi=L^{-(1+\frac{\nu}{\nu'})} \tau^{\frac{1}{z_u}}f(\tau L^{-z})$.

\begin{figure}[htbp]
 \includegraphics[width=0.5\columnwidth,clip]{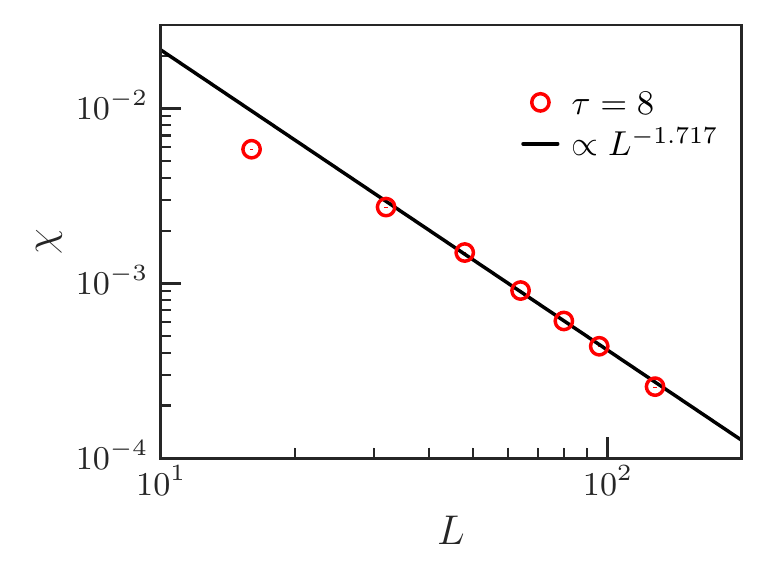}
  \vskip-3mm
  \caption{\textbf{Dependence of the susceptibility $\chi$ on $L$ for fixed $\tau$ in the short-time stage.} For $\tau=8$, $\chi$ scales with $L$ as $\chi\propto L^{-1.717}$ with the exponent being $1+\nu/\nu'$.}
  \label{fig:chil}
\end{figure}
However, it was shown that for larger system size the equilibrium scaling of $\chi$ crosses over to $\chi\propto L^{-\frac{\nu}{\nu'}}$~\cite{Shao2016sup}. This indicates that the above scaling form is not competent to describe the scaling behavior for large system sizes. We find that the full scaling form should be amended by including $\tau L^{-z_u}$ as an additional variable in the scaling function,
\begin{equation}
\label{eq:fullchiscaling}
\chi(\tau,L)=L^{-(d-z)}f(\tau L^{-z},\tau L^{-z_u}).
\end{equation}
For small system sizes, $\tau L^{-z}$ dominates and equation~(\ref{eq:fullchiscaling}) can be simplified as $\chi(\tau,L)=L^{-(d-z)}f(\tau L^{-z})\sim L^{-(d-z)}(\tau L^{-z})^{\frac{1}{z_u}}f(\tau L^{-z})$. For large system sizes, a possible crossover behavior is that equation~(\ref{eq:fullchiscaling}) can be approximated as $\chi(\tau,L)=L^{-(d-z)}f(\tau L^{-z},\tau L^{-z_u})= L^{-(d-z)}(\tau L^{-z})^{\frac{1}{z_u}}g(\tau L^{-z},\tau L^{-z_u})=L^{-\frac{\nu}{\nu'}}L^{1-(d-z)}h(\tau L^{-z},\tau L^{-z_u})$. For $\tau\rightarrow\infty$ and $d-z=1$, $g$ tends to a constant and thus $\chi\propto L^{-\frac{\nu}{\nu'}}$, while for small $\tau$, $\chi$ still satisfies $\chi\propto L^{-(1+\frac{\nu}{\nu'})}$. However, for the J-Q$_3$ model, numerical examination of this crossover behavior is beyond our present computational capacity, and we leave it for further studies.

\subsection{Critical initial slip}
In the main text, we shows that the time-correlation functions of the order parameters demonstrates a critical initial slip behavior with $\theta$ being its characteristic exponent. The scaling form of $C_P$ ($P$ represents $M$ or $D$) reads $C_P=\tau^{\theta}f_{C}(\tau L^{-z})$. In Fig.~\ref{fig:initialslip}, we further confirm this scaling form by showing that for fixed $\tau L^{-z}$, $C_P\propto \tau^{\theta}$ with $\theta\simeq0.514(3)$ for $P=M$ and $\theta\simeq0.546(7)$ for $P=D$.
\begin{figure}[htbp]
 \includegraphics[width=0.5\columnwidth,clip]{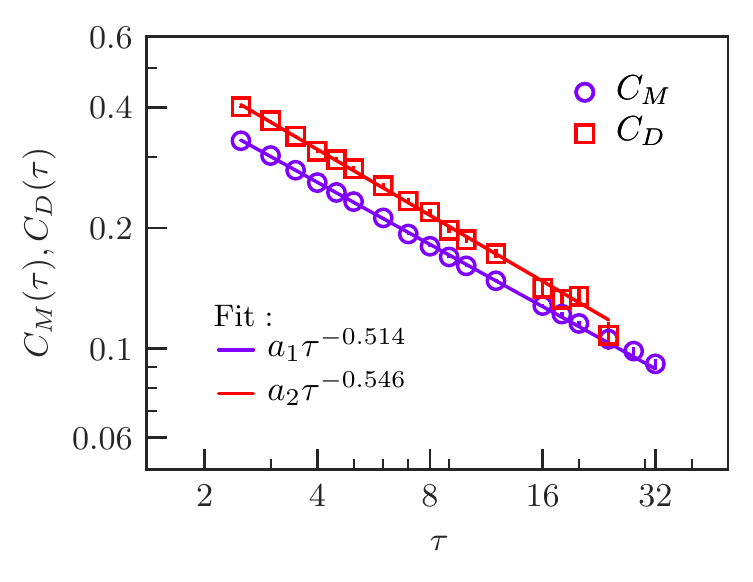}
  \vskip-3mm
  \caption{\textbf{Dependence of the time-correlation $C_M$ and $C_D$ on $\tau$ for fixed $\tau L^{-z}$.} For $\tau L^{-z}=1/4$, $C_M$ decays with $\tau$ as $C_M\propto \tau^{-0.514}$ and $C_D$ decays with $\tau$ as $C_D\propto \tau^{-0.546}$. These exponents are close to each other and they are just $\theta$.}
  \label{fig:initialslip1}
\end{figure}

\begin{figure}[htbp]
 \includegraphics[width=0.5\columnwidth,clip]{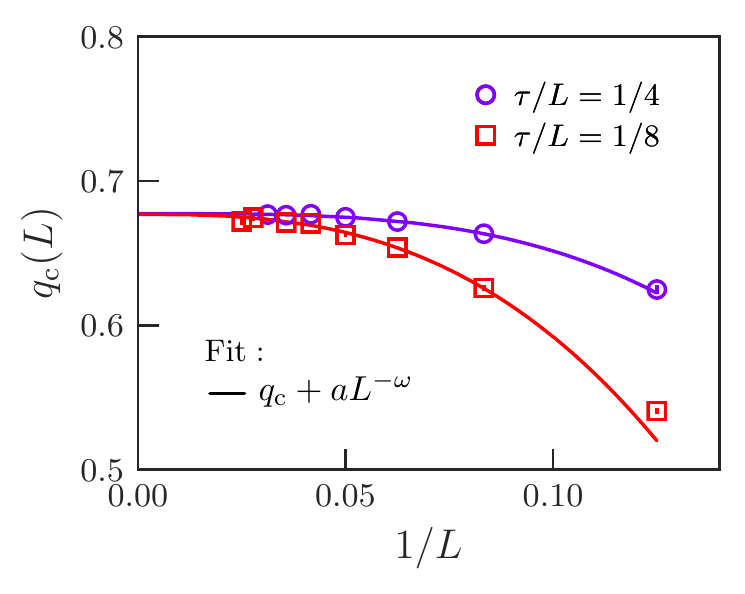}
  \vskip-3mm
  \caption{\textbf{Estimation of the critical from an unsaturated VBS state with $D_0=3/2L$.} For $\tau L^{-z}=1/8$, crossing points of $I_D-\delta$ for different $L$ converge onto $0.677(3)$. For $\tau L^{-z}=1/4$, crossing points of $I_D-\delta$ for different $L$ converge onto $0.6774(7)$. These results corroborate the value of critical point estimated in the main text.}
  \label{fig:escp}
\end{figure}

Moreover, the critical initial slip also manifests itself in the scaling behavior of the order parameter~\cite{Janssen1989sup,Yin2014sup,Shu2017sup,Shu2020sup}. In the usual LGW phase transitions, it was shown that in the short-time stage for an uncorrelated initial state with a small initial order parameter $P_0$, the order parameter $P$ changes as $P\sim P_0 \tau^\theta$. Scaling analyses show that $P_0$ has a dimension $x_0$ which is different from the dimension of $P$. Explicitly, $x_0=\theta z+\frac{\beta}{\nu}$. Directly generalizing this relation to the DQCP and comparing the exponent $\beta/\nu$ and $\theta$, one concludes that $x_0\simeq 0.1$. This small positive $x_0$ indicates that the initial order parameter $P_0$ is a weakly relevant scaling variable. Therefore, we infer that the dependence of some quantities on the initial order parameter is weak.

To examine this inference, we calculate $I_D$ for a different initial $D_0$. We find in Fig.~\ref{fig:escp} that for a fixed $\tau L^{-z}$ the crossing points of $I_D$ for different lattice sizes converge onto a point which is just the critical point within the error range, confirming that the dependence of the relaxation dynamics on the initial order parameter is weak.

Moreover, in the main text, we focus on the case when $P_0$ is saturated or zero. These two cases correspond to the fixed points of $P_0$. When $P_0$ is very small, $P$ changes as $P\sim P_0 \tau^\theta$; when $P_0$ is saturated, $P$ changes as $P\sim P_0 \tau^{-\beta/\nu z_u}$. Besides, in the usual LGW critical point, initial conditions with finite $P_0$ were also considered. In this case, the subsequent relaxation dynamics cannot be described by power functions. Instead, a complex universal characteristic function~\cite{Zheng1996sup,Zhang2014sup} was proposed. It is also instructive to study the universal characteristic function in the DQCP.

\clearpage

\section{Supplementary Information: Field Theory}
\def\bea{\begin{eqnarray}}
\def\eea{\end{eqnarray}}
\def\nn{\nonumber}
\def\ba{\begin{array}}
\def\ea{\end{array}}
\def\nn{\nonumber}
\def\Tr{\text{Tr}}
\def\sgn{\text{sgn}}
\def\J{\mathcal{J}}
\def\G{\mathcal{G}}
\def\O{\mathcal{O}}
\def\T{\mathcal{T}}
\def\R{\textcolor{red}}
\def\ii{\mathfrak{i}}
An equivalent description of the deconfined critical point in the J-Q model is given by SU$(2)$ QCD$_3$ where spinons in a fundamental representation interacts with SU$(2)$ gauge bosons in an adjoint representation~\cite{Tanaka2005sup, Senthil2006sup, Wang2017sup, Li2019sup}.
The lowest energy gauge potential configuration corresponds to the well-known $\pi$-flux in square lattice~\cite{Tanaka2005sup}.
To describe the low energy spinon moving in the background of $\pi$-flux square lattice, we define the following matrices as an explicit representation of Gamma matrices,
\bea
	\gamma^0 = \sigma^z, \gamma^1 =- \sigma^y \tau^z , \gamma^2 = -\sigma^x, \gamma^3= \sigma^y \tau^x, \gamma^5 = \sigma^x \tau^y,  \alpha^\mu = i \gamma^0 \gamma^\mu,
\eea
where $\bm \sigma$ and $\bm \tau$ act on sublattices and valleys, respectively.
The kinetic energy of spinon is given by
\bea \label{eq:kinetic}
	K = \psi^\dag \bm p \cdot \bm \alpha \psi, \quad \bm p = (p_1, p_y), \quad \bm \alpha = (\alpha^1, \alpha^2),
\eea
where $\psi$ ($\psi^\dag$) is the spinon annihilation (creation) operator, and we set the spinon velocity to one.
The matrices that correspond to SO$(5)$ orders are $(\alpha^3, \alpha^5, \gamma^0 \bm s)$, among which the first two are VBS orders and the last three are AFM orders with $\bm s$ acting on spins.
It is apparent that the kinetic energy is invariant under the SO$(5)$ rotation.

We consider the quench protocol that the initial spinon has a finite gap rendered by the background orders~\cite{Jian2019sup}.
We refer to VBS quench and AFM quench as the quench protocol where the initial state has a finite VBS and AFM order, respectively.
In general we need to consider individually the VBS and AFM quenches, 
however, as shown in the main text, a dual dynamic scaling appears as long as the universal low-energy dynamics is considered. So we can study one of them, and the other simply follows.
Let us consider the VBS quench without loss of generality. 

The prequench Hamiltonian is
\bea
	H = K + m \psi^\dag \alpha^5 \psi,
\eea
where $m$ characterizes the initial VBS order.
With the help of projection operator,
\bea \label{eq:prequench}
	P_\pm(m) = \frac12 \left(1 \pm ( \hat {\bm p}\cdot \bm \alpha +  \hat m \alpha^5) \right),
\eea
$\hat{\bm p} = (\sin \theta \cos \phi, \sin \theta \sin \phi)$, $\hat m = \cos\theta$, $\theta = \arctan \frac{m}{|\bm p|}$ and $\phi = \arctan \frac{p_y}{p_x}$, the solution for the prequnch Hamiltonian is
\bea
	\psi_i(\tau) = \left( e^{-\xi \tau} P_+(m) + e^{\xi \tau} P_-(m) \right) \psi_0, \quad \xi = \sqrt{\bm p^2 + m^2},
\eea
where $\psi_0$ is an arbitrary initial field configuration that will be integrated over.

At time zero, the finite order is suddenly turned off, $m=0$.
We first consider the free kinetic energy as the quench effect shows up, and then discuss the effect of interactions through $SU(2)$ gauge fields.
So in the lowest order without interactions, the postquench Hamiltonian is $K$ governing the dynamics on the Keldysh contour.

It is convenient to introduce the classical/quantum fields,
\bea
	\psi_{c/q} = \frac1{\sqrt2} (\psi_+ \pm \psi_-), \quad \psi_{c/q}^\dag = \frac1{\sqrt2} (\psi^\dag_+ \mp \psi^\dag_- ),
\eea
where $\psi_\pm$ are fields evolving forwards and backwards on the Keldysh contour.
And the boundary condition connected the Keldysh contour and the initial field is given by $\psi_+(0) =  \psi_i(\beta), \psi_-(0) = -\psi_i(0)$, where the minus sign is due to fermionic operators.
The propagator along the Keldysh contour is defined as
\bea
	i \hat G(t,t') \equiv  i \left( \ba{cccc}   G_R(t,t') & G_K(t,t') \\ 0 & G_A(t,t')  \ea \right) = \left( \ba{cccc}   \langle \psi_c(t) \psi^\dag_c(t') \rangle  & \langle \psi_c(t) \psi^\dag_q(t') \rangle \\ 0 & \langle \psi_q(t) \psi^\dag_q(t') \rangle \ea \right).
\eea
The retarded and advanced propagators are familiar,
\bea
	i G_R(t,t') = \theta(t-t') \left( e^{-i p (t-t')} P_+ + e^{i p (t-t')} P_- \right), \\
	i G_A(t,t') = -\theta(t'-t) \left( e^{-i p (t-t')} P_+ + e^{i p (t-t')} P_- \right).
\eea
with $p \equiv |\bm p|$ and $P_\pm \equiv P_\pm(0)$.
While the Keldysh propagator reads
\bea \label{eq:keldysh}
	i G_K( t,t') = \tanh \frac{\beta \xi}2 (e^{-i p t} P_+ + e^{i p t} P_-) \left(P_+(m) - P_-(m) \right)(e^{-i p t'} P_+ + e^{i p t'} P_-).
\eea
The VBS quench is encoded in the Keldysh propagator, and it is easy to check without the quench, $m=0$, (\ref{eq:keldysh}) restores time translation symmetry and reduces to equilibrium Keldysh propagator.

In order to calculate the equal time correlation function, we needs for instance the greater propagator
\bea
	G_+(t, t') \equiv -i \langle \psi_+(t) \psi_+(t') \rangle = \frac{1}2 \left(G_R(t,t') + G_A(t,t') + G_K(t,t')\right),
\eea
in which the subscript $+$ means the forward Keldysh contour. With these propagators, we now can evaluate the correlation function of different orders which are collected as follows,
\bea
	\bm \Phi = (\bm D, \bm M), \quad \bm D = (\psi^\dag \alpha^3 \psi, \psi^\dag \alpha^5 \psi), \quad \bm M = \psi^\dag \gamma^0 \bm s \psi.
\eea
Due to the $SO(5)$ symmetry, the correlation functions have two distinct types: one is parallel to the quenched order, and the other is orthonal to the quenched order, where the quenched order is $\Phi_2 = D_2$.

Before we calculate the equal time correlation function, we notice that the equal time correlator has a short distance singularity which is commonly seen in quantum field theory.
The order is a bilinear function of spinons, so it has the scaling dimension $[\bm D(\bm x, t)] = [\bm M(\bm x, t)] = D-1$, where $D$ is the space-time dimension.
This means the following equal time correlation functions
\bea
	D^2(t) &\equiv& \lim_{t' \rightarrow t^-} \int d^{D-1} \bm x \langle \bm D(\bm x, t) \cdot \bm D(0, t') \rangle, \\
	M^2(t) &\equiv& \lim_{t' \rightarrow t^-} \int d^{D-1} \bm x \langle \bm M(\bm x, t) \cdot \bm M(0, t') \rangle,
\eea
have divergence of $D-1$ order.
Here, since the system has spatial translational symmetry, one can utilize this freedom to choose the second order at origin $\bm x' = 0$.
It can be seen from the divergent part of the equilibrium correlation function
\bea \label{eq:UV}
	\lim_{t' \rightarrow t^-} \int d^{D-1} \bm x \langle \Phi_i(\bm x, t) \cdot \Phi_i(0, t') \rangle = 8 \int \frac{d^{D-1} \bm p}{(2\pi)^{D-1}},
\eea
which indeed diverges as $p^{D-1} $, and in our case $D=3$.
In order to cure this UV divergence, we simply subtract this divergent part, and we will use the same scheme in the quenched correlation function.

Let us look at the paralleled one,
\bea
	&&\int d^{2} \bm x \langle \Phi_2(\bm x, t)\Phi_2(0,t') \rangle = \int d^{2} \bm x \langle \psi^\dag_+(\bm x, t) \alpha^5 \psi_+(\bm x, t) \psi_+^\dag(0,t') \alpha^5 \psi_+(0,t') \rangle \\
	&=& m^2 + \int \frac{d^2\bm p}{(2\pi)^2} \Big[ \Tr\left( \alpha^5 G_R(\bm p, t, t') \alpha^5 G_A(\bm p, t', t) \right) + \Tr\left( \alpha^5 G_R(\bm p, t, t') \alpha^5 G_K(\bm p, t', t) \right) \\
	&& + \Tr\left( \alpha^5 G_K(\bm p, t, t') \alpha^5 G_A(\bm p, t', t) \right) + \Tr\left( \alpha^5 G_K(\bm p, t, t') \alpha^5 G_K(\bm p, t', t) \right) \Big]  \\
\label{eq:parallel}	&=& m^2 - \int \frac{d^2\bm p}{(2\pi)^2} \frac{8 m^2 \cos^2 (2 p t) }{p^2 + m^2 },
\eea
where we assume $t>t'$ and in the last step we take the limit $t' \rightarrow t^-$ and subtract the UV divergence (\ref{eq:UV}).
Notice that the first term is from the initial quenched order, and the second term starts to decrease the order since the postquench Hamiltonian locates at the critical point.

On the other hand, for the orthogonal one, $i \ne 2$, a similar calculation leads to
\bea
&&\int d^{2} \bm x \langle \Phi_i(\bm x, t)\Phi_i(0,t') \rangle = - \int \frac{d^2\bm p}{(2\pi)^2} \frac{8 m^2 \sin^2 (2 p t) }{p^2 + m^2 }, \label{eq:orthogonal}
\eea
Here we do not have a constant contribution since the initial order is zero.

It seems that the difference between $\cos^2$ and $\sin^2$ is irrelevant in (\ref{eq:parallel}) and (\ref{eq:orthogonal}).
However, we should remind readers that the actual simulation is the imaginary time projection where high energy modes are gradually projected out.
As a crude approximation, we can try to capture this effect by multiplying a Boltzmann weight $e^{- p |t|}$ which effectively project out the high-energy modes $p> \frac1{|t|}$.
In this situation, the difference between (\ref{eq:parallel}) and (\ref{eq:orthogonal}) is apparent since the integral mainly concentrate at $ p |t| \ll 1$, so the parallel correlation function (\ref{eq:parallel}) receives a big contribution from quenched order whereas the orthogonal one (\ref{eq:orthogonal}) is intact from the quenched order!

Now let us discuss two length (time) scales at the critical point.
First notice that the emergent SU$(2)$ QCD$_3$ theory features Lorentz symmetry, which sets the dynamical critical exponent to one $z=1$.
Although the quench protocol explicitly breaks the Lorentz symmetry, we assume the critical dynamical exponent remains unchanged the fixed point.

A hallmark of the deconfined quantum critical point is the dynamics of deconfinement of gauge fields, and this brings a new length scale---the confinement length $\xi'$---that is distinct from the conventional correlation length scale.
More explicitly, in the language of deconfined spinon, the confinement length $\xi'$ is associated with the  Wilson loop
\bea
	W(x,y) = \langle \bar \psi (x) e^{i \int_{x}^y d x'^\mu A_\mu(x')} \psi(y) \rangle \propto e^{- \frac{|x-y|}{\xi'}},
\eea
where $A$ denotes the $SU(2)$ gauge field, whereas the conventional correlation length scale is associated with the spinon bilinear mass, i.e., $G(x,y) \sim e^{-|x-y|/\xi}$.


Physically, the confinement length $\xi'$ is the width of VBS domain wall or equivalently the spinon confinement length~\cite{Nahum2015sup,Shao2016sup}. So it is greater than the conventional correlation length $\xi$, and these two diverge at different scalings,
\bea
	\xi \propto \delta^{-\nu}, \quad \xi' \propto \delta^{-\nu'},
\eea
where $ \delta = q-q_c $ is the distance to the critical point, and $\nu'>\nu$.

For the VBS quench, the parallel (VBS) correlation function in (\ref{eq:parallel}) remembers the finite initial order which enters as the correlation length $\xi \sim 1/m$. In this case $\xi \propto \tau^{1/z_u}$, $z_u \equiv z \nu'/\nu $ dominates and $D^2$ behaves as
\bea	\label{scalingd}
	D^2(\tau, L) &=& L^{- 2\beta/\nu} F_D( \tau/L^{\nu/\nu'}),
\eea
This equation can be transformed as
\bea	\label{scalingd}
	D^2(\tau, L) &=& \tau^{- 2(\beta/\nu)(\nu/\nu') } F_{D1}( \tau/L^{\nu/\nu'}).
\eea
This is just the scaling form of $D^2$ discussed in the main text. 
In the short-time stage, $\tau\ll L^z$, $D^2(\tau, L)$ is mainly affected by the initial condition but almost independent of $L$. In this case, $D^2(\tau, L)\propto \tau^{- 2(\beta/\nu)(\nu/\nu')}=\tau^{- 2(\beta/\nu z_u)}$.

Note that here one may argue that $D^2$ can also be expressed as $D^2(\tau, L)\propto \tau^{- 2\beta/\nu' z}$. In this case, $2\beta/\nu'\simeq 0.881$ is smaller than one. Accordingly, the scaling law $2\beta/\nu'-1=\eta'$ gives a negative anomalous dimension $\eta'$, which would imply a nonunitary theory~\cite{Nahum2015sup,Ferrara1974sup,Mack1977sup}. 
Alternatively, to satisfy the unitarity bound~\cite{Ferrara1974sup,Mack1977sup}, we choose to adopt $D^2(\tau, L)\propto \tau^{- 2\beta/\nu z_u}$ in which ${2 \beta/\nu}$ keeps intact while an additional dynamic exponent $z_u$ is introduced. We want to emphasize that the appearance of this $z_u$ does not mean that the equilibrium correlation time needs two dynamic exponents. This $z_u$ only measures the scaling relation between the age of the system $\tau$ and the correlation length $\xi$.

For the orthogonal (AFM) correlation in (\ref{eq:orthogonal}) the effect of the initial order is washed away in the imaginary evolution but the intrinsic deconfined dynamics with length scale $\xi' \propto \tau^{1/z}$ is intact.
Thus the correct scaling function resulted from VBS quench should be
\bea \label{eq:orthogonal_scale}
	M^2(\tau, L) &=& L^{- 2\beta/\nu} F_M( \tau/L),
\eea
When $\tau\ll L^z$, $M^2$ behaves as $L^{-d}$ according to the central limit theorem. Combining with the scaling requirements that $M^2\propto \tau^{-2\beta/\nu z}$, one obtain $M^2\propto L^{-d}\tau^{d/z-2\beta/\nu z}$ in the short-time stage. This is just the scaling form of $M^2$ discussed in the main text.
Moreover, if one starts from a random disordered state, it is natural to expect that the initial quench protocol does not contain any useful information that can be probed by any simple observables.
Therefore, both the VBS and AFM orders will have similar behaviors like the orthogonal correlation where the intrinsic deconfinement length scale $\xi' \propto \tau^{1/z}$ takes place and leads to (\ref{eq:orthogonal_scale}).
This is fully consistent with the simulation starting from a disordered state in the main text.

\end{widetext}

\end{document}